\documentclass[prd,twocolumn,a4paper,superscriptaddress,floatfix]{revtex4-1}
\usepackage{graphicx}
\usepackage{amssymb}
\usepackage{amsfonts}
\usepackage{amsmath}
\usepackage{color}
\usepackage{ulem}
\usepackage[utf8]{inputenc}
\usepackage[T1]{fontenc}

\begin{document}

\title{Non-minimal coupling contribution to DIS at low $x$ in Holographic QCD}

\author{Artur Amorim}
\affiliation{Centro de F\'{\i}sica do Porto e Departamento de F\'{\i}sica e Astronomia da Faculdade de Ci\^encias da Universidade do Porto, Rua do Campo Alegre 687, 4169-007 Porto, Portugal}
\author{Robert Carcass\'{e}s Quevedo}
\affiliation{Centro de F\'{\i}sica do Porto e Departamento de F\'{\i}sica e Astronomia da Faculdade de Ci\^encias da Universidade do Porto, Rua do Campo Alegre 687, 4169-007 Porto, Portugal}
\author{Miguel S. Costa}
\affiliation{Centro de F\'{\i}sica do Porto e Departamento de F\'{\i}sica e Astronomia da Faculdade de Ci\^encias da Universidade do Porto, Rua do Campo Alegre 687, 4169-007 Porto, Portugal}

\begin{abstract}
We consider the effect of including a non-minimal coupling between a $U(1)$ vector gauge field and the graviton Regge trajectory in holographic QCD models. This coupling describes the QCD interaction between the quark bilinear electromagnetic current and the Pomeron. We test this new 
coupling against DIS data at low Bjorken $x$ and obtain an excellent fit with a chi squared of 1.1 over a very large kinematical range in the photon virtuality $Q^2<400 \ {\rm GeV}^2$ and for  $x<10^{-2}$. 
The scale of the new dimension full coupling, which arises from integrating  higher spin fields,  is of order  $6\  {\rm GeV}$. 
This value matches precisely the expectations from effective field theory, which indicate that such corrections are controlled by the mass gap between
the spin two and spin four glueballs that are described holographically by the graviton and spin four field in the graviton Regge trajectory, respectively
\end{abstract}
\maketitle

%%%%%%%%%%%%%%%%%%%%%%%%%%%%%%%%%%%
\section{\label{sec:Int} Introduction}
%%%%%%%%%%%%%%%%%%%%%%%%%%%%%%%%%%%

The observation that the Pomeron is dual to the graviton Regge trajectory \cite{brower_pomeron_2007} opened an entirely new approach
to the analysis of QCD processes dominated by Pomeron exchange. 
This fact has been explored in diffractive processes, like low-$x$ deep inelastic scattering (DIS) \cite{BallonBayona:2007qr,hatta_deep_2008,cornalba_saturation_2008,pire_ads/qcd_2008,albacete_dis_2008,hatta_relating_2008,levin_glauber-gribov_2009,brower_saturation_2008,brower_elastic_2009,gao_polarized_2009,hatta_polarized_2009,kovchegov_comparing_2009,avsar_shockwaves_2009,cornalba_deep_2010,dominguez_particle_2010,cornalba_ads_2010,betemps_diffractive_2010,gao_polarized_2010,kovchegov_$r$_2010,levin_inelastic_2010,domokos_pomeron_2009,domokos_setting_2010,brower_string-gauge_2010,ballon_bayona_unity_2017}, deeply virtual Compton scattering \cite{costa_deeply_2012}, vector meson production \cite{costa_vector_2013}, double diffractive Higgs production \cite{Brower:2012mk}, central production of mesons \cite{anderson_central_2014} and other inclusive processes \cite{Nally:2017nsp}. It is now clear that 
holographic QCD is a valuable tool to model the physics of gluon rich medium, where  standard perturbative techniques like the BFKL pomeron
 \cite{Fadin:1975cb,Kuraev:1977fs,Balitsky:1978ic} breakdown.

In this paper we focus on low $x$ DIS, extending the previous work \cite{ballon_bayona_unity_2017}. The basic idea is to 
construct the holographic Regge theory for the glueball exchange associated with the Pomeron trajectory. 
In DIS the Pomeron couples to the quark bilinear electromagnetic current $J^\mu=\bar{\psi} \gamma^\mu \psi$, which 
is described holographically  by the 
interaction between a bulk $U(1)$ vector gauge field and the graviton Regge trajectory. Here we shall extend the analysis of 
 \cite{ballon_bayona_unity_2017} by allowing for a non-minimal coupling between this gauge field and the higher spin 
 fields in the graviton Regge trajectory. We shall fit the same set of data as in  \cite{ballon_bayona_unity_2017}, more concretely we fit 
 249 data points, covering the very large kinematical range of $x<10^{-2}$ and $Q^2<400 \ {\rm GeV}^2$, where $x$ is the Bjorken $x$ and $Q^2$ the photon virtuality. As a result, we manage to improve the quality of our fit from a chi squared per degree of freedom of 1.7 in  \cite{ballon_bayona_unity_2017} to an excellent value of 1.1 in the present work. 

The existence of such non-minimal coupling between the bulk $U(1)$ gauge field and the graviton Regge trajectory is expected. 
Starting from the UV high energy limit, the OPE  expansion of the two currents, $J_\mu(x) J_\nu(y)$, contains two OPE coefficients for each spin $J$ symmetric traceless operator associated with the glueballs on the pomeron trajectory, 
${\cal O}_J \sim {\rm tr}(F_{\mu\alpha_1} D_{\alpha_2}\cdots D_{\alpha_{J-1}} F_{\alpha_J}^{\ \ \mu}) $.
Holographically, and for pure AdS space,  this amounts to precisely the same counting when coupling 
a vector gauge field to the graviton, or to the higher spin fields in the gravity Regge trajectory.
Thus we shall consider such non-minimal coupling. In fact, since QCD is not a conformal theory, there is actually more freedom in the 
choice of such couplings in holographic QCD which, as we shall see, are very much model dependent. 
For concreteness we shall consider one such coupling, which arises in an effective field theory expansion in the
dual QCD string tension. After obtaining the new expression for the DIS structure function $F_2(x,Q^2)$ in generic
AdS/QCD models, we focus on the specify holographic QCD model of  \cite{gursoy_exploring_2008,gursoy_exploring_2008-1,gursoy_improved_2011}.
This allows us to put numbers in our expressions that are then tested against available low $x$ DIS data.

%%%%%%%%%%%%%%%%%%%%%%%%%%%%%%%%%%%
\section{Holographic computation of $F_2$ structure function}
%%%%%%%%%%%%%%%%%%%%%%%%%%%%%%%%%%%

The structure function $F_2(x,Q^2)$ is related to the total cross-section of the inelastic  $\gamma^*p\to X$  process. 
As discuss in the standard literature (see for instance~\cite{devenish_deep_2004}),
defining $\sigma_T$ and $\sigma_L$ to be the cross sections for 
transverse and longitudinal polarizations, we have
\begin{equation}
 \sigma_T + \sigma_L = \frac{4\pi^2\alpha} {Q^2}\, F_2(x,Q^2)\,,
 \label{eq:sigma}
\end{equation}
where  $\alpha$ is the fine structure constant.
The structure function depends on the 
photon virtuality $Q^2$ and on the Bjorken $x\ll1$, which we take to be small. 
Through the optical theorem, this total cross-section can be related to the imaginary part of the amplitude ${\cal A}$ for  elastic forward scattering 
$\gamma^*p\to\gamma^*p$, with the appropriate polarizations. Thus
\begin{equation}
  \label{eq:F2 from A}
  F_2(x,Q^2)=\frac{Q^2}{4\pi^2\alpha} \,\frac{1}{s}\, \text{Im}\,\mathcal{A}(s,t=0)\,,
\end{equation}
where $s$ and $t$ are the usual Mandelstam variables (in the low $x$ regime, $s=Q^2/x$). 
We will compute this amplitude using the AdS/QCD prescription as described below.

First let us define our kinematic variables. We use light-cone coordinates $\left(+,-,\perp \right)$, with the flat space metric given by $ds^2 = - dx^+ dx^- + d x^2_\perp$, where $x_\perp \in \mathbb{R}^2$ is a vector in impact parameter space. 
We take for the large $s$ kinematics of  $12\to34$ scattering the following
\begin{align}
  \label{eq:kinematics}
&k_1=\left(\!\sqrt{s},-\frac{Q^2}{\sqrt{s}} ,0\right),\  \ k_3=-\left(\!\sqrt{s},\frac{ q_\perp^2 -Q^2}{\sqrt{s}} , q_\perp \right)\!,\\
&k_2=\left(\frac{M^2}{\sqrt{s}},\sqrt{s} ,0\right),\  \ k_4=-\left(\frac{M^2+ q_\perp^2}{\sqrt{s}},\sqrt{s} ,-q_\perp \right).
\nonumber
\end{align}
where $k_1$ and $k_3$ are  respectively the incoming  and outgoing photon momenta. The proton 
target has mass $M$ and incoming  and outgoing  momenta $k_2$ and $k_4$, respectively.
For the forward scattering considered in the optical theorem we set $q_\perp = 0$, so that $k_1=-k_3$, and we take the same polarization
for the  incoming and outgoing photon. The possible polarization vectors are
\begin{equation}
  \label{eq:polarization vectors} 
  n(\lambda)=\begin{cases}
    (0,0,\epsilon_\lambda) \,, & \lambda=1,2 \,,\\
    \left( \sqrt{s}/Q, Q/\sqrt{s},0 \right) , & \lambda=3\,,
    \end{cases}
\end{equation}
where $\epsilon_\lambda$ is just the usual transverse polarization vector.

%%%%%%%%%%%%%%%%%%%%%%%%%%%%%%%%%%%
\subsection{AdS/QCD}
%%%%%%%%%%%%%%%%%%%%%%%%%%%%%%%%%%%

\begin{figure}[t!]
  \center
  \includegraphics[height=4cm]{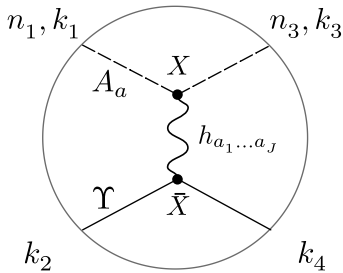} 
  \caption{Tree level Witten diagram representing spin $J$    exchange in a $12\to34$ scattering. 
%     Double lines are used to denote graviton's propagator. The $n_1$ and $n_3$ labels denote incoming/outgoing proton's polarizations respectively, for the forward scattering $n_1=n_3$.
}
  \label{fig:Witten_diagram}
\end{figure}

We shall compute the above scattering amplitude using the framework of AdS/QCD. First we present general formulae and then specify 
to a particular model. As explained in the introduction, we are interested in the Regge limit where the amplitude is dominated by the exchange 
of the graviton Regge trajectory, which includes fields of even spin $J$. We also need to define our holographic external states. 
The corresponding Witten diagram is shown in figure~\ref{fig:Witten_diagram}. The upper part of the diagram is related to the incoming and outgoing virtual photons, whereas the bottom part to the proton target.

The holographic dual of QCD will have a dilaton field and a five-dimensional metric, which in the vacuum will have the form
\begin{equation}
ds^2 = e^{2 A(z)} \left[ dz^2 + \eta_{\mu \nu} dx^\mu dx^\nu \right]    \,, \qquad
\Phi = \Phi(z) \,, \label{IHQCDBackground}
\end{equation}
for some unknown functions $A(z)$ and $\Phi(z)$. 
The dilaton is dual to the Lagrangian and the metric to the  energy-momentum tensor.
We shall use greek indices in the boundary, with flat metric $ \eta_{\mu \nu}$.
We will work with the string frame metric.

In DIS the external photon is a source for the conserved $U(1)$ current $\bar{\psi} \gamma^\mu \psi$,
where the quark field $\psi$ is associated to the open string sector. 
The five dimensional dual of this current is a massless $U(1)$ gauge field $A$. We shall assume that this 
field is made out of open strings and that is non-minimally coupled to the metric, with the following action
\begin{equation} 
  S_A = - \frac{1}{4} \int d^5 X \sqrt{-g} \, e^{-\Phi} \left( F_{ab}F^{ab}  + \beta R_{a b c d}F^{ab}F^{cd} \right),
\label{eq:u1 action}
\end{equation}
where $F=dA$ and we use the notation $X^a=(z,x^\alpha)$ for five-dimensional points. 
The corresponding equation of motion can be easily derived to be
\begin{equation}
  \label{eq:u1 eom}
  \nabla_a \left[ e^{-\Phi}\left( F^{ab} + {\beta} R^{ab}_{\, \, \, \, \, cd}F^{cd} \right)\right]=0\,.
\end{equation}
The coupling $\beta$ has dimensions of length squared. At this order in derivatives of the fields, we could have other couplings 
to the Riemann tensor, to derivatives of the dilaton field and also higher derivative terms in the field strength $F$. As we shall see bellow, we will be mostly interested in the coupling to the graviton in the linearised theory, in which case there are only two possible local couplings. Thus, for our purposes  the above action is rather general.

We will fix the gauge of the $U(1)$ bulk field to be $D_a A^a = 0$, which gives  
$A_z = 0$ and $\partial_\mu A^\mu = 0$. 
The solution of the equation of motion (\ref{eq:u1 eom}) in this gauge is then
\begin{equation}
 A_\mu^\lambda \left( X ; k\right) =  n_\mu^\lambda \, f_k ( z )\,e^{i k \cdot x}\,,
\end{equation}
where $f_k(z)$ solves the differential equation
\begin{equation}
  \label{eq:u1 eom gauge fixed}
  \left[-Q^2+e^{\Phi-A}\partial_z\left(e^{A-\Phi}\partial_z \right) +\beta \Delta_{\beta} \right]f_Q(z)= 0 \,,
\end{equation}
with
\begin{equation}
\Delta_{\beta}=-2 e^{-2A}\left[\left(-\dot A \ddot A - \dot \Phi \ddot A + \dddot A\right)\partial_z + \ddot A \partial_z^2 - \dot A^2 Q^2 \right] .
\end{equation}
Notice that here, and in the remainder of this paper, we shall denote derivatives with respect to $z$ with a dot.
The momentum  $k$ and the polarisation vector $n^\lambda$ satisfy
\begin{equation}
  k^2 = Q^2 \,, \qquad 
  n^\lambda_z = 0 \,, \qquad
   k \cdot n^\lambda = 0 \,,
\end{equation}
where the boundary polarisation is given by~\ref{eq:polarization vectors}. 
We choose as UV boundary condition $f(0)=1$ which gives the non-normalizable solution,
since the off-shell photon acts as a source for the quark bilinear current $\bar{\psi} \gamma^\mu \psi$.
Finally, let us note that, for the computation of the Witten diagram in figure~\ref{fig:Witten_diagram},
it is convenient to compute the field strength of a given mode 
\begin{align}
 F_{\mu\nu} (X;k,n) & = 2 i k_{[\mu}n_{\nu]}f_{Q}(z)e^{i k\cdot x} \,, \notag\\
  F_{z\mu} (X;k,n)   &  = n_{\mu}  \dot{f}_{Q}(z)e^{i k\cdot x} \,,
  \label{eq:U1 F_ab}
\end{align}
where $Q^2=k^2$.

For the proton target we consider a scalar field $\Upsilon$ that represents an unpolarised proton described by a normalizable mode of the form
\begin{equation}
\Upsilon(X;p )= \upsilon_m(z) \, e^{i p\cdot x} \,,
\label{eq:proton}
\end{equation}
where $p$ is the momentum and $m^2=-p^2$.
As explained in detail in~\cite{ballon_bayona_unity_2017},
the specific details of the function will not be important because it will appear in an integral 
that can be absorbed in the coupling between the pomeron and the proton. 

%%%%%%%%%%%%%%%%%%%%%%%%%%%%%%%%%%%
\subsection{Non-minimal coupling}
%%%%%%%%%%%%%%%%%%%%%%%%%%%%%%%%%%%

To compute the Witten diagram of  figure~\ref{fig:Witten_diagram}, we need 
to consider the interaction between the external scattering states and the 
spin $J$ fields in the graviton Regge trajectory. Thus, the higher spin field comes from the closed string sector 
while the external fields come from the open sector. 

First we consider the coupling between the $U(1)$ gauge field and the graviton. In Einstein-Maxwell theory, and for $AdS$ or flat space, it is well known
that there are only two possible cubic couplings between these fields, namely
\begin{equation}
F^{ac}F^{b}_{\ c}h_{ab} \,, \quad 
 F^{ac}F^{bd}\nabla_{c} \nabla_{d} h_{ab}\,,
 \label{eq:two_couplings}
\end{equation}
where $h_{ab}$ is the metric fluctuation.
The present case, however, is less restrictive because we have an additional 
scalar field and also because  space-time is not maximally symmetric. 
To understand this better, let us  linearize the  action  (\ref{eq:u1 action}) around the background  metric, that is, we write $g_{ab}=\bar{g}_{ab}+h_{ab}$. Setting $h=h^a_{\ a}=0$ we have the cubic couplings
%In the gauge $D_a h^{ab}=0$ and $h^a_{\ a}=0$  we have the cubic couplings
 \begin{align}
  \delta S &= - \frac{1}{2} \int d^5 X \sqrt{-\bar{g}} \,e^{-\Phi} \Big(  F^{ab} F^{c}_{\ b} h_{a c}   
    \label{eq:linearized_action}
  \\   & \qquad
    +\frac{\beta}{2} h_{a p} \bar{R}^{p}_{b c d} F^{a b}F^{cd} -  \beta  F^{a c} F^{b d} \bar{\nabla}_a \bar{\nabla}_b h_{c d}\Big)\,.
  \nonumber
 \end{align}
To study the graviton Regge trajectory in the background (\ref{IHQCDBackground}) we need to 
decompose the metric in $SO(1,3)$ irreducible representations. We will be only interested in the 
graviton TT components $h_{\alpha\beta}$, satisfying $\partial^\alpha h_{\alpha\beta}=0$ and 
$h^{\alpha}_{\ \alpha}=0$, and we set $h_{z\alpha}=0=h_{zz}$.
Using that $R_{\alpha\mu\beta\nu}= \dot{A}^2 e^{2A} (\eta_{\alpha\nu}\eta_{\mu\beta}-\eta_{\alpha\beta}\eta_{\mu\nu}  )$ and
$R_{\alpha z\beta z}= -\ddot{A} e^{2A} \eta_{\alpha\beta}$ in the background (\ref{IHQCDBackground}), and computing the covariant derivatives,
we obtain
\begin{align}
   \delta S &= - \frac{1}{2} \int d^5 X \sqrt{-\bar{g}} \,e^{-\Phi} \Big[ F^{\alpha \mu} F^{\beta}_{\ \mu} \big(1 - \beta e^{-2A} \dot{A} \partial_z \big) 
  \nonumber\\
  &  - \beta F^{\alpha \mu} F^{\beta \nu} \partial_\mu \partial_\nu 
  - 2 \beta F^{\alpha z} F^{\beta\nu} \big( \partial_z - 2 \dot{A} \big) \partial_\nu
\label{eq:linearized_action2}\\
   & +F^{\alpha z} F^{\beta}_{\ z} \left(1 - \beta e^{-2A} \big( \partial_z^2 - 3 \dot{A} \partial_z + 2 \dot{A}^2 \big)  \right) \Big] h_{\alpha\beta}\,.
    \nonumber
\end{align}
Notice that in the AdS case ($A=-\log z$) these couplings reduce to the two allowed couplings in (\ref{eq:two_couplings}).
However, in the present case there are more possibilities. For example, other contractions with the Riemann tensor will give
 different functions multiplying the 
same tensor structures in the couplings. We may also use derivatives of the scalar field to contract with the field strength. For simplicity, 
the approach we follow in this work will be to focus on the coupling given by the action  (\ref{eq:u1 action}). Our aim is to test whether this type of corrections are important in describing DIS using holographic QCD.

Next we wish to generalize the previous coupling to case of the cubic interaction between the gauge field and a symmetric, transverse and traceless spin $J$ field,  $h_{a_1\dots a_J}$. The pomeron trajectory includes such higher spin fields of even $J$.
Again there are several possibilities, but we shall focus on the simplest extension of the two couplings to the graviton considered above. 
The first term is the minimal coupling term, which can be generalized to 
\begin{eqnarray}
  \label{spin_J_MC_U1}
  \kappa_J \!
   \int \! d^5 X \sqrt{-g} \, e^{-\Phi }  F^{a_1 b} \bar{\nabla}^{a_2} \dots \bar{\nabla}^{a_{J-1}}F^{a_J}_{\ \, b} h_{a_1\dots a_J} .
  \label{eq:couplingA}
\end{eqnarray}
The transverse condition of $h_{a_1\dots a_J}$ guarantees that this term is unique up to dilaton derivatives. For the non-minimal coupling we will write
\begin{align}
  \label{spin_J_NMC_U1}
  & \beta_J \int d^{d+1} X \sqrt{-g} e^{-\Phi}\Big(  F^{ca_1}\bar{\nabla}^{a_2}\dots\bar{\nabla}^{a_{J-1}}F^{a_J d }\bar{\nabla}_c\bar{\nabla}_d  
 \notag \\
  & \qquad + \frac{1}{2} F^{a_1 b} \bar{\nabla}^{a_2}\dots \bar{\nabla}^{a_{J-1}}F^{cd} R^{a_J}_{\ \ b c d}  \Big)h_{a_1 \dots a_J} \,.
\end{align}
We remark that in both expressions (\ref{eq:couplingA}) and (\ref{spin_J_NMC_U1}) the way we distribute the covariant derivatives acting on the field strength is important. After integrating by parts such a covariant derivative, we are left with an extra term in the derivative of the background dilaton field. However, these terms will have a component of the higher spin field along the $z$ direction, which can be dropped in the case of 
the pomeron.

Next we need to 
decompose the spin $J$ fields in $SO(1,3)$ irreducible representations. In the Regge limit
we are only interested in the TT components of these fields, that is in $h_{\alpha_1 \dots \alpha_J}$ with
$\partial^{\nu}h_{\nu \alpha_2 \dots \alpha_J}=0$ and $h^\nu_{\ \nu\alpha_3 \dots \alpha_J}=0$. From now on we will
assume these two conditions. Thus for  the minimal coupling  (\ref{eq:couplingA}) we obtain simply
\begin{align}
  \kappa_J  \int  d^5 X \sqrt{-g} \, e^{-\Phi } \Big( F^{\alpha_1 \mu} \partial^{\alpha_2} \dots \partial^{\alpha_{J-1}}F^{\alpha_J}_{\ \, \mu}
  \nonumber\\
+  F^{\alpha_1 z} \partial^{\alpha_2} \dots \partial^{\alpha_{J-1}}F^{\alpha_J}_{\ \, z}  \Big)
  h_{\alpha_1\dots \alpha_J} .
\end{align}
For the non-minimal coupling (\ref{spin_J_NMC_U1}) we obtain after a cumbersome computation
\begin{align}
  &\beta_J \int d^{5} X \sqrt{-g} e^{-\Phi} \Big[  F^{ z\alpha_1 } \partial^{\alpha_2}\cdots\partial^{\alpha_{J-1}}F^{\alpha_J}_{\ \  z}{\cal D}^J_\parallel +
  \notag \\ 
  &    F^{\mu \alpha_1} \partial^{\alpha_2}\cdots \partial^{\alpha_{J-1}}F^{\alpha_J \nu} 
  \left( e^{2A} {\cal D}^J_\perp  \eta_{\mu\nu} + \partial_\mu \partial_\nu \right) +
   \label{eq:nmcoupling}\\ 
     \notag
  & \left. 2 F^{ \mu \alpha_1 } \partial^{\alpha_2} \cdots \partial^{\alpha_{J-1}} F^{\alpha_J z} \left( \partial_z-J \dot{A}  \right) \partial_\mu 
  \right] h_{\alpha_1 \cdots \alpha_J} \,,
\end{align}
where
\begin{align}
     {\cal D}^J_\perp =&\ e^{-2A} \dot{A} \left( \partial_z - \left( J -2 \right) \dot{A} \right)  , \notag \\
     {\cal D}^J_\parallel = &\ e^{-2A} \left( \partial_z^2 - \left( 2 J -1 \right) \dot{A} \partial_z \right. 
     \label{eq:Ds} \\
    &  \left.  - \left( J - 2 \right) \ddot{A} + J \left( J - 1 \right){\dot{A}}^2 \right)\notag .
\end{align}
For $J=2$ this coupling reduces to the graviton non-minimal coupling given in  (\ref{eq:linearized_action2}).

For the scalar field $\Upsilon$ we will consider a minimal coupling with spin J closed string fields
\begin{eqnarray}
\bar{\kappa}_J \int d^5 X \sqrt{-g} \, e^{-\Phi } \, \left ( \Upsilon \nabla^{a_1} \dots \nabla^{a_J} \Upsilon \right ) \,  h_{a_1 \dots a_J} \, .
\label{eq:couplingScalar}
\end{eqnarray}
Again, this coupling is unique up to derivatives of the dilaton field that are subleading in the Regge limit. Focusing on the TT part of the spin $J$
field, we are left with the single coupling
\begin{eqnarray}
\bar{\kappa}_J \int d^5 X \sqrt{-g} \, e^{-\Phi } \, \left ( \Upsilon \partial^{\alpha_1} \dots \partial^{\alpha_J} \Upsilon \right ) \,  h_{\alpha_1 \dots \alpha_J} \, .
\label{eq:couplingScalar2}
\end{eqnarray}

%%%%%%%%%%%%%%%%%%%%%%%%%%%%%%%%%%%
\subsection{Witten diagram in Regge limit}
%%%%%%%%%%%%%%%%%%%%%%%%%%%%%%%%%%%

The scattering amplitude will have a contribution from the minimal and the non-minimal coupling. The contribution of the minimal coupling to the structure function $F_2$ is presented and described in \cite{ballon_bayona_unity_2017}.
Here we shall compute the contribution of the non-minimal coupling (\ref{eq:nmcoupling}) to the exchange of a spin $J$ field, 
corresponding to  the 
Witten  diagram in figure  \ref{fig:Witten_diagram}. Using the Regge kinematics  (\ref{eq:kinematics}) and taking as external states  $F_i^{ab}(X)$ for $i=1,3$ and $\Upsilon_j(\bar{X})$ for $j=2,4$, we obtain for forward scattering the expression
\begin{align}
  & \beta_J  \bar{\kappa}_J \sum_{\lambda = 1}^3 \int d^5 X d^5 \bar{X} \sqrt{-g} \sqrt{-\bar{g}} \, e^{-\Phi} e^{- \bar{\Phi}}
 \Upsilon_{2} {\left( \bar{\partial}^{-} \right)}^J \Upsilon_{4}   \notag\\ 
&\left[ F_{1}^{\,+z} \left( \partial^+\right)^{J-2} F_{3\,\, z}^{\,+} {\cal D}^J_\parallel  +
F_{1}^{\,+\mu} \left( \partial^+\right)^{J-2} F_{3\,\, \mu}^{\,+}  {\cal D}^J_\perp  \right] \\
 & \times \Pi_{+\cdots +,-\cdots -}(X, \bar{X} )\,,
  \notag
\end{align}
where bars denote quantities evaluated at $\bar{X}$.
Notice that the couplings involving derivatives along the boundary in (\ref{eq:nmcoupling}) vanish for forward scattering.
Using  (\ref{eq:U1 F_ab})  and  (\ref{eq:proton}) for the external states and performing the sum over polarisations we find
\begin{align}
     & -  \beta_J \bar{\kappa}_J s^J \int d^5 X d^5 \bar{X} \sqrt{-g} \sqrt{-\bar{g}} \,e^{-\Phi- \bar{\Phi} -2(J+1)A -2J \bar{A}}  
     \notag \\
     &  \times \upsilon^2_m (\bar{z })     \left( f_Q^2(z) {\cal D}^J_{\perp} + \frac{{\dot{f}_Q^2(z)}}{Q^2} {\cal D}^J_\parallel \right) 
     \Pi_{+ \cdots +, - \cdots -} \,  .
     \end{align}
 We remark that the terms with  
 $\mathcal{D}_\perp^J$ and with $\mathcal{D}_\parallel^J$ are, respectively, the  leading contribution arising from the transverse and longitudinal polarizations, therefore justifying our notation.

By changing variable $w = x - \bar{x}$ and defining the transverse propagator at zero momentum transfer by 
\begin{align}
    & \int \frac{d w^{+} d w^{-} d^2 l_\perp}{2}\, \Pi_{+ \cdots +, - \cdots -} \left( w, z, \bar{z}\right) = \\ \notag
    & = - \frac{i}{2^J}  e^{ (J-1) (A + \bar{A})}    G_J ( z, \bar{z}, t=0),
\end{align}
we finally obtain
\begin{align}
    & i \frac{ \beta_J \bar{\kappa}_J  s^J}{2^J} V \int dz d\bar{z} e^{-\Phi - \bar{\Phi} -2J (A +\bar{A}) +3A +5\bar{A} }\upsilon^2_m(\bar{z})
 \\ 
    &      \times  \left( f_Q^2(z) {\cal D}^J_\perp + \frac{{\dot{f}}_Q^2(z)}{Q^2} {\cal D}^J_\parallel \right) \left[ e^{ (J-1) (A + \bar{A})}G_J (z, \bar{z}, 0) \right] .
 \notag
 \end{align}
Now we proceed as in~\cite{ballon_bayona_unity_2017} and write a spectral representation for the transverse propagator 
\begin{align}
    G_J ( z, \bar{z}, t ) = e^{B + \bar{B}} \sum_n \frac{\psi_n (J, z ) \, \psi^{*}_n (J, \bar{z})}{t_n( J ) - t}\,,
    \label{eq:TransProp}
\end{align}
where $\psi_n (J, z )$ are the normalizable modes associated to the spin $J$ fields.
The function $B(z)$ depends on the particular holographic QCD model. We will fix it later in order to perform fits to data.

%%%%%%%%%%%%%%%%%%%%%%%%%%%%%%%%%%%%%%%%%%%%%
\subsection{Regge Theory}
%%%%%%%%%%%%%%%%%%%%%%%%%%%%%%%%%%%%%%%%%%%%%

In order to get the total amplitude we need to sum over even spin J fields with $J \geq 2$. Then we can apply a Sommerfeld-Watson transform
\begin{align}
  \frac{1}{2} \sum_{J \geq 2} \left( s^J + (- s)^J \right)= - \frac{\pi}{2} \int \frac{dJ}{2 \pi i} \frac{s^J + (- s)^J}{\sin \pi J}\,,
\end{align}
which requires  analytic continuation of the amplitude for spin $J$ exchange to the complex J-plane.
We assume that the J-plane integral can be deformed from the poles at even J, to the poles $J = j_n (t)$ defined by $t_n (J) = t$. The scattering domain of negative t contains these poles along the real axis for $J < 2$. The scattering amplitude for $t=0$ is then
\begin{align}
  & A(s,0) = \sum_n h_n s^{j_n} \int dz \,e^{-\Phi} e^{A(-2 j_n + 3)} \times  \\
  & \left( f_Q^2 {\cal D}^{j_n( 0)}_\perp + \frac{{\dot{f}}_Q^2}{Q^2} {\cal D}^{j_n ( 0 )}_\parallel \right) 
  \left[ e^{A(j_n ( 0)-1)} e^{B} \psi_n \big(j_n ( 0 ), z \big)  \right] ,
  \notag
\end{align}
with $h_n$ defined as
\begin{align}
  & h_n = - \frac{\pi}{2} \frac{{\beta}_{j_n(0)}\bar{\kappa}_{j_n(0)}}{2^{j_n(0)}} \left( i +  \cot\frac{\pi j_n (0)}{2} \right) j'_n (0)  \\ 
  & \times \int d \bar{z}\, e^{\bar{A}(4-j_n (0))} e^{-\bar{\Phi}} e^{\bar{B}} {\upsilon^2_m ( \bar{z})} \, \psi^{*}_n \big(j_n ( 0 ), \bar{z} \big) \,.
  \label{eq:h_n}
\end{align}

Finally, the action of the differential operators on the functions of $z$ allows us to rewrite the forward scattering amplitude as
\begin{align}
  & A(s,0) = \sum_n h_n s^{j_n} \int dz\, e^{-(j - 2) A + B - \Phi}  \notag \\ 
  & \times \left( f_Q^2 \tilde{{\cal D}}^{j_n \left( 0 \right)}_\perp + \frac{{\dot{f}}_Q^2}{Q^2} \tilde{{\cal D}}^{j_n \left( 0 \right)}_\parallel \right) 
  \psi_n \big( j_n ( 0 ), z  \big) \,,
  \label{eq:FinalForward}
\end{align}
with
\begin{align}
   \tilde{{\cal D}}_\perp = \ &e^{-2A} \left(\dot{A} \partial_z + \dot{A}^2 + \dot{A}\dot{B}\right) ,  \\
   \tilde{{\cal D}}_\parallel = \ &e^{-2A} \left( \partial_z^2 - \big( \dot{A} - 2 \dot{B} \big) \partial_z
  + \ddot{B} + \ddot{A} + \dot{B}^2 - \dot{A}\dot{B} \right).
   \notag
\end{align}

%%%%%%%%%%%%%%%%%%%%%%%%%%%%%%%%%%%%%%%%%%%%%
\subsection{$F_2$ structure function}
%%%%%%%%%%%%%%%%%%%%%%%%%%%%%%%%%%%%%%%%%%%%%

The DIS structure function can be written in Regge theory in the following form
\begin{equation}
  F_2(x, Q^2) = \sum_{n}   \Big(  f^{\text{MC}}_{n}(Q^2) + f^{\text{NMC}}_{n}(Q^2)  \Big) x^{1-j_n}\,,
\end{equation}
where  we separated the contributions from the minimal and non-minimal couplings between the 
graviton trajectory and the $U(1)$ current that arise from the holographic computation. In \cite{ballon_bayona_unity_2017} we showed that 
 \begin{equation} 
  f^{\text{MC}}_{n}(Q^2)  = g_n Q^{2 j_n} \int dz \,e^{-\left(j_n-\frac{3}{2}\right)A}  \left( f_Q^2  +  \frac{\dot{f}_Q^{2}}{Q^2}      \right) \psi_n\,.
  \label{eq: f_n MC} 
\end{equation}
Using the definitions (\ref{eq:sigma}) and (\ref{eq:F2 from A}), we may 
 take the imaginary part of the forward scattering (\ref{eq:FinalForward}), to obtain
the contribution from the non-minimal coupling
 \begin{align} 
  f^{\text{NMC}}_{n}(Q^2)  =\ & \tilde g_n Q^{2 j_n} \int dz \,e^{-\left(j_n-\frac{3}{2}\right)A} \times \nonumber\\
  \label{eq: f_n NMC} 
  &  \left( f_Q^2 \mathcal{\tilde{\cal D}}_\perp +  \frac{\dot{f}_Q^{2}}{Q^2} \mathcal{\tilde{\cal D}}_\parallel
     \right) \psi_n\,,
\end{align}
where $\tilde g_n={\rm Im} (h_n)/(4\pi^2\alpha)$. Both constants $g_n$ and $\tilde g_n$    are used  as fitting parameters in our setup, thus the details of holographic wave function for the proton are not important in the fit.
Notice that the $g_n$ and $\tilde g_n$  do not have the same dimensions, indeed comparing both complings we see that 
$[\tilde g_n/g_n]=L^2$.
Formula (\ref{eq: f_n NMC}) is one of the main results of this paper.

 % together with the parameters of the potential of the Schrodinger problem for the $\psi_n(z)$, are the input of our model. 
 
 %Let us notice that the $x$ and $Q^2$ dependence already has the form suggested by Donnachie and Landshoff in~\cite{donnachie_small_1998,donnachie_new_2001}. The dependence of $F_2$ in $x$ is just a power law whereas the dependence in $Q^2$ is more complicated since $Q$ appears as a parameter in the integrals over $z$. In the next section we will present our result for the set of parameters which explains better the experimental data.

%%%%%%%%%%%%%%%%%%%%%%%%%%%%%%%%%%%%%%%%%%%%%
\subsection{Improved Holographic QCD}
%%%%%%%%%%%%%%%%%%%%%%%%%%%%%%%%%%%%%%%%%%%%%

To test the above ideas against experimental data we need to consider a concrete QCD holographic model. As in our previous work 
\cite{ballon_bayona_unity_2017}, we shall consider the improved holographic QCD model introduced in \cite{gursoy_exploring_2008,gursoy_exploring_2008-1,gursoy_improved_2011}. This fixes the
background fields  $A(z)$ and $\Phi(z)$, which give an approximate dual description of the QCD vacuum.

Next we need to consider the equation of motion for the spin $J$ fields that are dual to the twist two operators, whose exchange gives the dominate 
contribution in DIS at low $x$. This equation is then analytically continue in $J$, in order to do the  Sommerfeld-Watson transform in Regge theory. 
This procedure was described in detail in \cite{ballon_bayona_unity_2017}, so we will not repeat it here.  The upshot is that 
the function $B$ introduced in (\ref{eq:TransProp}) to define the transverse propagator is given by $B=\Phi - A/2$ and 
the normalisable modes of the spin $J$ field $\psi_n(z)$ solve a Schr\"{o}dinger problem
\begin{equation*}
  \left(-\frac{d^2}{dz^2}+U_J(z)\right)\psi_n(z)=t_n\psi_n(z)\,,
\end{equation*}
where
\begin{align*}
  U_J(z)=&\  \frac{3}{2}\left(\ddot A - \frac{2}{3}\ddot \Phi\right) + \frac{9}{4}{\left(\dot A - \frac{2}{3}\dot \Phi \right)}^2\\
  & + (J - 2)e^{-2A}\bigg[\frac{2}{l^2_s}\left(1+\frac{d}{\sqrt{\lambda}}\right) + \frac{J + 2}{\lambda^{4/3}}\\
  &+e^{2A} \left(a \ddot \Phi + b \left(\ddot A - \dot{A}^2\right) + c \dot{\Phi}^2  \right)\bigg]\,,
\end{align*}
where the first line  represents the potential for the graviton and the remaining proposed terms deform the graviton potential.
This potential is analytically continued in $J$ in such a way that  the value of the intercept $J=j_n$ 
is obtained when the $n$-th  eigenvalue satisfies $t_n(J)=0$.

The constants $l_s$, $a$, $b$, $c$ and $d$ are used as fitting parameters and will be adjusted such that the best match with $F_2(x,Q^2)$ data is achieved. In particular, from the low energy effective string theory perspective, $l_s$ is related to the string tension; $d$ is related to the anomalous dimension curve of the twist 2 operators, or it can also be thought as encoding the information of how the masses of the closed strings excitations are corrected in a slightly curved background; the constants $a$, $b$ and $c$ encode the first order derivative expansion of a presumed string field theory lagrangian.

%%%%%%%%%%%%%%%%%%%%%%%%%%%%%%%%%%%
\section{\label{sec:Data} Data analysis}
%%%%%%%%%%%%%%%%%%%%%%%%%%%%%%%%%%%

\begin{figure}[t!]
  \center
  \hspace*{-1.5cm}
  \includegraphics[height=10cm]{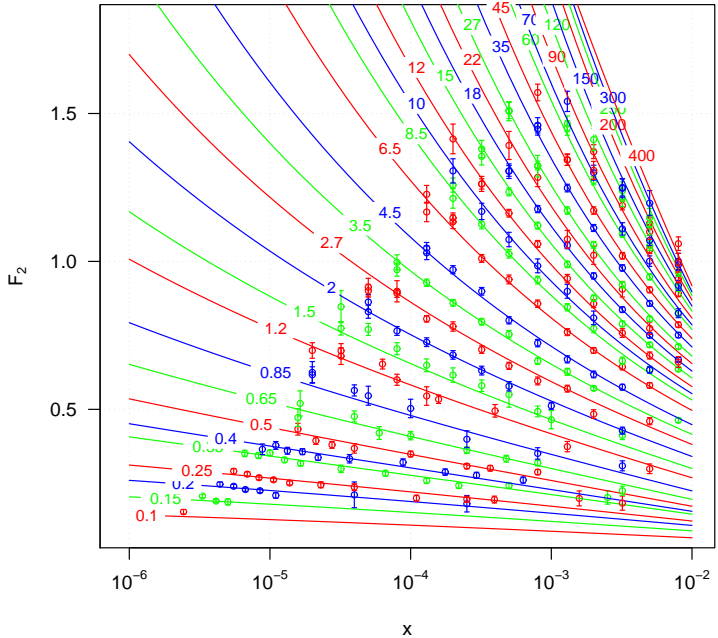} 
  \caption{Structure function $F_2(Q^2,x)$. Experimental points vs prediction of this work with a  $\chi_{d.o.f}^2=1.1$. Each line
  corresponds to a given $Q^2 \ ({\rm GeV}^2)$ as indicated.}
\label{fig:F2}
\end{figure}

\begin{figure}[t!]
  \center
  \includegraphics[height=9cm]{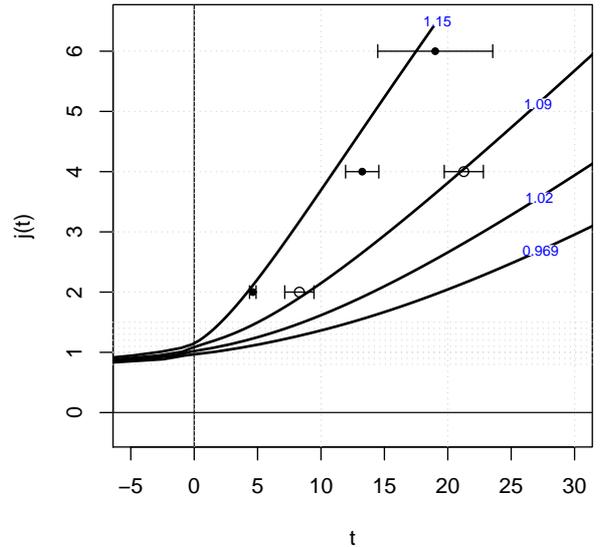} 
  \caption{Regge trajectories compared with glueball masses from lattice simulations \cite{meyer_glueball_2005,meyer_glueball_2005-1}.
  Shown are also the values we obtained for the intercept of each trajectory. 
  Configurations that give the soft pomeron intercept $j_1=1.09$ were favoured in the fitting process.}
\label{fig:trajectories}
\end{figure}

With the previously described setup we proceed to find the best values for the potential parameters $l_s$, $a$, $b$, $c$ and $d$, as well as for the coupling values $\beta$, $g_n$ and $\tilde {g}_n$ that better fit the data. We look, as usual, for the best set of parameter values such that the sum of the weighted difference squared between experimental data and model predicted values is minimum, using as weight the inverse of the experimental uncertainty. Since this is a highly non trivial numerical optimization problem in which we do not known explicitly the gradient of the function to be optimized, we use the Nelder-Mead algorithm, using $R$ language, and try with different starting points in the parameter space. We have found that the inclusion of the non-minimal coupling contribution considerable decreases the convergence ratio of the minimizing routine compared with the case where only the minimal coupling case is used, consistent with the fact that the new function to optimize has a much rougher landscape. Our best fit results for $F_2(x,Q^2)$ are presented in figure~\ref{fig:F2}. In this fit we considered values of $x$ in the range $x<10^{-2}$, and of the photon virtuality $Q^2<400 \ {\rm GeV}^2$. This gives a total number of $249$ data points.
The $\chi^2_{d.o.f}$ for this fit is $1.13$. As in our previous work, aiming to make a consistent model for the Soft Pomeron, we have forced the intercept of the second trajectory to be around $j_1=1.09$. This is achieved penalizing those set of parameters which give a different second intercept by adding a term of the type $10^4(j_1-1.09)^2$ to the function to be optimized.
The correspondent Regge trajectories can be seen in figure~\ref{fig:trajectories}.

The values of the parameters that give the best fit are summarized in table~\ref{table:best fit parameters}. 
We would like to understand the scale defined by the non-minimal coupling. The best fit fixes the 
value of this coupling in the equation of motion (\ref{eq:u1 eom}) for the $U(1)$ gauge field to be 
$\beta=0.026\  {\rm GeV}^{-2}$. Thus the energy scale associated with this correction is about $6 \ {\rm GeV}$.
Alternatively we may look
at the ratio between the constants $g_n$ and $\tilde {g}_n$, given by, 
\begin{equation}
\frac{\tilde {g}_n}{g_n}= \frac{\beta_{j_n(0)}}{\kappa_{j_n(0)}}\,,
\end{equation}
which has dimensions $length{^2}$.
This follows from taking the imaginary part of (\ref{eq:h_n}) and from the fact that ${g}_n$ has a similar expression.
 Looking at table~\ref{table:best fit parameters} we see that the analytic continuation of the 
 non-minimal coupling is also at the same energy scale. This scale 
 should be associated with the  mass gap between the spin 2 and spin 4 glueballs, that arise from the spectrum of the 
 bulk graviton and spin 4 field, respectively. Indeed this is precisely the size of the gap 
 observed in the glueball spectrum in figure \ref{fig:trajectories}.

%In particular we have found that $\alpha=-0.0672$. 

\begin{table}[b!]
  \centering
  \caption{Values of the parameters for the best fit found.
  All parameters are dimensionless except for $[l_s]=L$, $[\beta]=L^2$ and $[\tilde{g}_i]=L^2$.
  Numerical values are expressed in ${\rm GeV}$ units.}
\label{table:best fit parameters}
  \begin{tabular}{|c|c|c|c|c|c|}
  \hline
  parameter & value  & couplings   & value     &  couplings   & value $\times 10$      \\ \hline
  $l_s^{-1}$  & 6.93   & $g_0$  & -0.154    & $\tilde g_0$ & 0.707                \\ \hline
  a           & -4.68  & $g_1$  & -0.424    & $\tilde g_1$ &  -0.378              \\ \hline
  b           & 4.85   & $g_2$  & 2.12      & $\tilde g_2$ &  -2.48              \\ \hline
  c           & 0.665  & $g_3$  & -0.721    & $\tilde g_3$ &  3.63               \\ \hline
  d           & -0.328 &  &    &            &                                \\ \hline
  $\beta$ & -0.026 &  &  &              &                        \\ \hline
  
  \end{tabular}
  \end{table}

%%%%%%%%%%%%%%%%%%%%%%%%%%%%%%%%%%%
\section{\label{sec:Conclusions} Conclusion}
%%%%%%%%%%%%%%%%%%%%%%%%%%%%%%%%%%%

\begin{figure}[t!]
  \center
  \includegraphics[height=9cm]{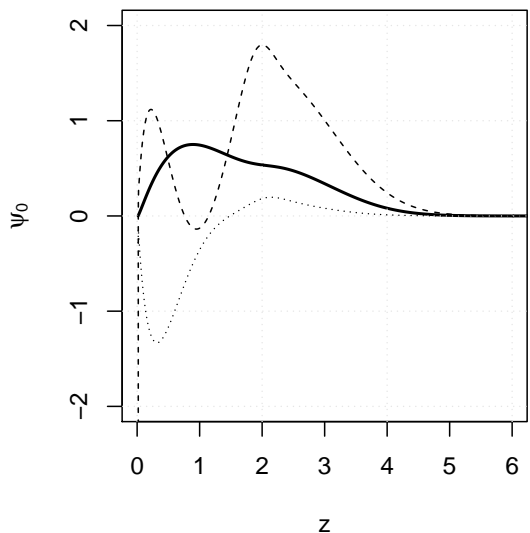} 
  \caption{Hard Pomeron wave function for the best fit found and for its intercept value $J=j_0$. The dotted and dashed line represent the action of the operator $\mathcal{D}_\perp$ and $\mathcal{D}_\parallel$ on the Hard Pomeron wave function $\psi_0(z)$ respectively. In this plot all the functions have been scaled by a factor of $10$.}
\label{fig:Psi0_vs_Dperp_Dparallel_Psi}
\end{figure}

In this work we  considered the contribution of a non-minimal coupling between the $U(1)$ gauge field and the higher spin fields in the  graviton Regge trajectory to the holographic computation of the DIS structure function $F_2(x,Q^2)$. These non-minimal couplings are expected to be present and to play an important role in theories with higher spin fields. Such terms are  controlled by the gap between the graviton and the next higher spin field \cite{camanho_causality_2016}. Our results are consisten with this expectation since the scale we obtained for the non-minimal coupling has the correct order of magnitude that reproduces the mass difference between the spin 2 and spin 4 glueballs.

With the inclusion of the new coupling the quality of our fit to low $x$ DIS data has improved considerably. In the previous work \cite{ballon_bayona_unity_2017}, that considered only the minimal coupling, a 
$\chi^2_{d.o.f}$  of 1.7 was obtained. With the new coupling we improved this result to a $\chi^2_{d.o.f}$  of 1.1. We believe this is
an important improvement that validates the holographic approach to low $x$ physics. We are reproducing data over a very large kinematical range in the two variables $x$  and $Q^2$, fitting a total of 249 points.
 
 \begin{figure}[t!]
  \center
  \includegraphics[height=9cm]{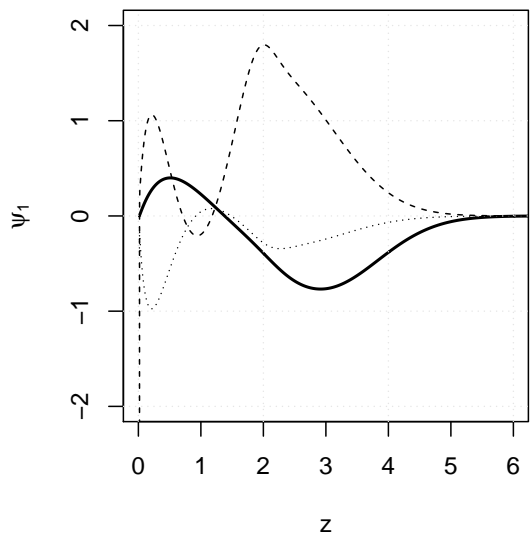} 
  \caption{Same as figure~\ref{fig:Psi0_vs_Dperp_Dparallel_Psi} but for the Soft Pomeron and for its intercept value 
  $J=j_1$.}
\label{fig:Psi1_vs_Dperp_Dparallel_Psi}
\end{figure}

One can draw some intuition on how the inclusion of the non-minimal coupling improves the fit to physical data by looking at the Reggeon wave functions. These functions are shown for the hard and soft pomerons, for the corresponding values of the intercept,  in figures \ref{fig:Psi0_vs_Dperp_Dparallel_Psi} and \ref{fig:Psi1_vs_Dperp_Dparallel_Psi}, respectively. These waves functions are the 
ground state and first excited state of the associated Schr\"{o}dinger problem. For the minimal coupling they control the dependence of the structure function in the photon virtuality $Q^2$ as can be seen from (\ref{eq: f_n MC}).  For the non-minimal coupling they also control the $Q^2$ dependence 
but now the action of the differential operators $\tilde{{\cal D}}_\perp$ and $\tilde{{\cal D}}_\parallel$  in  (\ref{eq: f_n NMC}) changes such dependence to a more oscillating  behaviour, as can be seen from figures \ref{fig:Psi0_vs_Dperp_Dparallel_Psi} and \ref{fig:Psi1_vs_Dperp_Dparallel_Psi}.  
What is not a priori trivial is that this freedom can be used to better fit the data, yielding  for the scale of non-minimal coupling 
precisely the expected order of magnitude (due to the oscillations it could be that this order of magnitude was much smaller, which would seem to contradict the expected value of the gap for  higher spin glueballs).

It seems we are getting closer to a very satisfactory holographic description of low $x$ data. There are two immediate questions that we believe deserve some further attention. As a working example we have been considering the improved holographic QCD model of 
\cite{gursoy_exploring_2008,gursoy_exploring_2008-1,gursoy_improved_2011}.  We take this model as our QCD vacuum, and then introduce higher spins fields for which we do Regge theory. Clearly we should study to which extent other models can also be used to reproduce the data here analysed. Our expectation is that holography is very appropriate to study processes dominated by Pomeron exchange, so that other models that are close enough to QCD should give similar results. Another interesting point is to extend this analysis to other processes than DIS. Previous studies of deeply virtual Compton scattering (DVSC) and vector meson production could now be revisited, including the non-minimal coupling here considered, to attain better fits. For example, in the case of DVSC the cross section depends on three kinematical quantities, namely $x$, $Q^2$ and momentum transfer. Extending the contribution of the non-minimal coupling terms  to non-vanishing $t$ gives a very non-trivial dependence that deserves to be looked at.

\section{Acknowledgments}
We would like to thank A. Ballon Bayona for useful discussions. 
This research received funding from the grant CERN/FIS-PAR/0019/2017 and the 
Simons Foundation grant 488637 (Simons collaboration on the Non-perturbative bootstrap). 
AA is funded by the IDPASC doctorate programme with the fellowship 
PD/BD/114158/2016.

%%%%%%%%%%%%
%%%%%%%%%%%%
%%%%%%%%%%%%
\appendix
\bibliography{bib/pomeron,%
              bib/refs,%
              bib/AdSCFT,%
              bib/glueballs,%
              bib/HQCD}

\end{document}